\documentclass[apjl]{emulateapj}
\usepackage{amsfonts,amsmath,amssymb}
\usepackage{graphicx}
\usepackage{hyperref}
\usepackage{color}
\usepackage{url}
\usepackage{multirow}

\usepackage{txfonts}
\usepackage[dvipsnames,usenames]{xcolor}

\newcommand{\code}[1]{\texttt{#1}}

\newcommand{\mesa}{\code{MESA}}

\newcommand{\nuclei}[2]{\mathrm{^{#1}#2}}

\newcommand{\helium}[1][4]{\nuclei{#1}{He}}

\newcommand{\beryllium}[1][9]{\nuclei{#1}{Be}}

\newcommand{\carbon}[1][12]{\nuclei{#1}{C}}

\newcommand{\silicon}[1][28]{\nuclei{#1}{Si}}

\newcommand{\Ledd}{L_{\mathrm{Edd}}} 
\newcommand{\Tc}{T_{\mathrm{\!c}}} 
\newcommand{\Mdot}{\dot{M}} 

\newcommand{\ee}[1]{\times 10^{#1}}		
\newcommand{\power}[2]{{#1^{#2}}}		    

\newcommand{\unitskip}{\,}                             

\newcommand{\unitstyle}[1]{\mathrm{#1}}

\newcommand{\Mega}{\unitstyle{M}}		

\newcommand{\Kelvin}{\unitstyle{K}}
\newcommand{\K}{\Kelvin}  

\newcommand{\yr}{\unitstyle{yr}}               

\newcommand{\Msun}{M_\odot}
\newcommand{\Lsun}{L_\odot}

\newcommand{\Myr}{\Mega\yr}

\newcommand{\Msunperyr}{\Msun\unitskip\power{\yr}{-1}}	

\newcommand{\Mign}{M_\mathrm{ign}}
\newcommand{\Lnuc}{L_\mathrm{nuc}}

\begin{document}
\title{Direct Three-body Triple-$\alpha$ in Helium Novae}
\shorttitle{DIRECT THREE-BODY TRIPLE-$\alpha$ IN HELIUM NOVAE}
\author{Ryan Connolly\altaffilmark{1,2}, Alex Deibel\altaffilmark{1,2}, and Edward F. Brown\altaffilmark{1,2,3}}
\affil{
\altaffilmark{1}{Department of Physics and Astronomy, Michigan State University, East Lansing, MI 48824, USA: connol65@msu.edu} \\
\altaffilmark{2}{The Joint Institute for Nuclear Astrophysics - Center for the Evolution of the Elements, Michigan State University, East Lansing, MI 48824, USA} \\
\altaffilmark{3}{National Superconducting Cyclotron Laboratory, Michigan State University,
East Lansing, MI 48824, USA} 
}

\shortauthors{CONNOLLY, DEIBEL, \& BROWN}


\begin{abstract}
In AM CVn binaries, a white dwarf primary accretes material from a helium-rich white dwarf or stellar companion. The unstable ignition of nuclear burning via the $3\alpha$ reaction in an accumulated helium layer powers a thermonuclear runaway near accretion rates $\dot{M} \lesssim 10^{-6} \, \mathrm{M_{\odot} \ yr^{-1}}$ that may be observed as helium nova or .Ia supernova. Helium burning in the primary's envelope at temperatures $T \lesssim 10^{8} \, \mathrm{K}$ may proceed via the direct three-body fusion of $\alpha$-particles. Here we show that the direct three-body rate by \citet{Nguyen_2012} \textrm{---} which is reduced relative to the extrapolated resonant rate at temperatures $T \gtrsim 5 \times 10^{7} \, \mathrm{K}$ \textrm{---} results in novae with longer recurrence times and larger ignition masses. By contrast, we find that the enhancement in the direct three-body rate at temperatures below $T \lesssim 5 \times 10^{7} \, \mathrm{K}$ does not result in significant differences in nova outburst properties. The most massive envelopes in our models are near the density threshold for detonation of the helium layer, where an increase in the density at ignition due to the $3\alpha$ rate may be important.
\end{abstract}

\keywords{binaries: close \textrm{---} novae, supernovae: general \textrm{---} white dwarfs}

\section{Introduction} \label{sec:intro}

In ultra-compact AM Canum Venaticorum (AM CVn) binaries, a He or CO white dwarf (hereafter WD) primary accretes helium-rich material from a low-mass companion \citep{Warner_1995}. Accretion onto the WD surface near temperatures $T \sim 10^{7}\textrm{--}10^{8} \, \mathrm{K}$ may lead to unstable ignition of the helium layer resulting in a helium nova, analogous to classical novae with hydrogen. In the case of a degenerate donor star, mass transfer increases the orbital separation of the binary and the accretion rate consequently decreases with time \citep{Deloye_2003, Deloye_2005, Bildsten_2006}. As the accretion rate drops, subsequent ignitions require a thicker helium shell, resulting in increasingly energetic novae. The final and most energetic of these events are proposed as ``.Ia supernovae" \citep{Bildsten_2007}. This mechanism has been discussed for several underluminous supernovae \citep{2008ApJ...683L..29K,2009AJ....138..376F,Poznanski_2010,Kasliwal_2010} and \citet{Bildsten_2007} predict that hundreds may be observed per year with deep nightly surveys such as the Large Synoptic Survey Telescope.

At temperatures $T \gtrsim 10^{8} \, \mathrm{K}$, the $3\alpha$ burning of accumulated helium in the primary's envelope proceeds sequentially via the reaction $\alpha + \alpha \rightleftharpoons \beryllium[8]$ followed by $\beryllium[8] + \alpha \rightarrow \carbon + \gamma$ \citep{Hoyle_1954}. Stellar evolution models typically use extrapolations (e.g., NACRE; \citealt{Angulo_1999}) of the resonant rate for $\beryllium[8] + \alpha \rightarrow \carbon + \gamma$ to lower temperatures and neglect non-resonant contributions. At temperatures $T \lesssim 10^8\,\mathrm{K}$, however, helium burning may occur through the direct three-body fusion of $\alpha$-particles. \citet{Ogata_2009} calculated a direct three-body $3\alpha$ rate at these temperatures that resulted in a substantial enhancement below $T \lesssim 2 \times 10^8\, \K$ compared to the NACRE rate. \citet{Dotter_2009} showed that the enhanced rate, however, had dire implications for the evolution of low mass stars; core helium ignition occurred much sooner, shortening or even eliminating the red giant branch. \citet{Nguyen_2012} calculated a direct three-body 3$\alpha$ rate using a hyperspherical harmonics R-matrix method (hereafter HHR). Although the HHR rate is enhanced relative to NACRE at temperatures $T \lesssim 5 \times 10^7\, \K$ (and reduced at temperatures $T \gtrsim 5 \times 10^{7} \, \mathrm{K}$), stellar evolution calculations are consistent with the results using the NACRE rate \citep{Nguyen_2012}. 

Here we investigate the HHR 3$\alpha$ rate by \citet{Nguyen_2012} in AM CVn systems and its impact on helium novae. The unstable ignition of helium is sensitive to the $3\alpha$ rate at temperatures $T \lesssim 10^{8} \, \mathrm{K}$ where the HHR rate differs significantly from the NACRE rate. We outline our accreting WD models in Section~\ref{sec:methods} and describe the accretion history for different types of donor stars. The type of donor determines the evolution of the accretion rate over time, which in turn affects the properties and frequency of the helium novae, especially the most energetic events. Section \ref{sec:results} details the results of our simulations and the effects of modifying the $3\alpha$ rate on the properties of the helium novae. In particular, we will show that the HHR rate results in larger helium ignition masses and longer recurrence times for helium novae. We discuss our results in Section \ref{sec:conclusions}.

\section{White dwarf model and the 3$\alpha$ rate}\label{sec:methods}

We calculate accretion onto a carbon-oxygen (CO) white dwarf using with the open-source stellar evolution code \mesa \ \citep{Paxton_2011,Paxton_2013,Paxton_2015}. 
For radiative opacities, we use the ``Type 2" OPAL opacities that account for enhancements of C and O \citep{Iglesias_1993,Iglesias_1996} supplemented by \citet{Ferguson_2005} at low temperatures. We use the $\code{co\_burn}$ reaction network that includes hydrogen, helium, carbon, and oxygen burning, as well as $\alpha$-chain reactions up to $\silicon[28]$. Accretion is shut off during most novae, specifically when the nuclear burning luminosity $\Lnuc \gg \Ledd$, and resumes when $\Lnuc \lesssim \Ledd$. Note that for a $1 \, \Msun$ WD the Eddington luminosity is $\Ledd  \approx  3 \times 10^{4} \, L_{\odot}$.

\begin{figure}
\centering
\includegraphics[width=1.0\columnwidth]{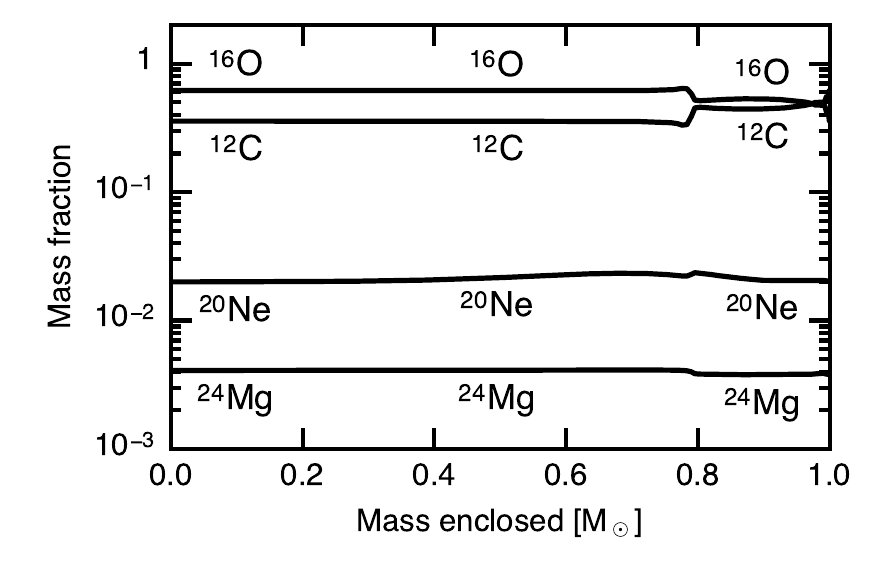}
\vspace{-0.4cm}
\caption{Initial composition as a function of enclosed mass for our CO WD model. There is no envelope on the WD initially; helium accretes directly onto the CO core without processing any existing material.}
\label{fig:composition}
\end{figure}

In tight binaries such as AM CVn systems, mass loss via Roche lobe overflow is important as the envelope expands during a nova \citep{Piersanti_2015} and may result in a merger of the component stars during a common envelope event \citep{2015ApJ...805L...6S}. We choose not to include mass loss by Roche lobe overflow, however, and instead the model uses a super-Eddington wind scheme \citep[see][]{Denissenkov_2013}. This mass loss prescription should be adequate for the study here because we want to focus on the ignition conditions of each nova rather than the evolution of the novae themselves. The wind serves only to shed most of the envelope during each nova. Figure~\ref{fig:composition} shows the initial composition of the CO WD prior to accretion. The H-rich envelope has been stripped from the WD core and the model neglects any processing of a pre-existing envelope due to compressional heating by accretion. Note that an initial hydrogen flash would pre-heat the envelope and could change the properties of the first few helium novae, but this behavior is transient and likely negligible in the long-term evolution of the model. 

\begin{figure} 
\centering
\includegraphics[width=1.0\columnwidth]{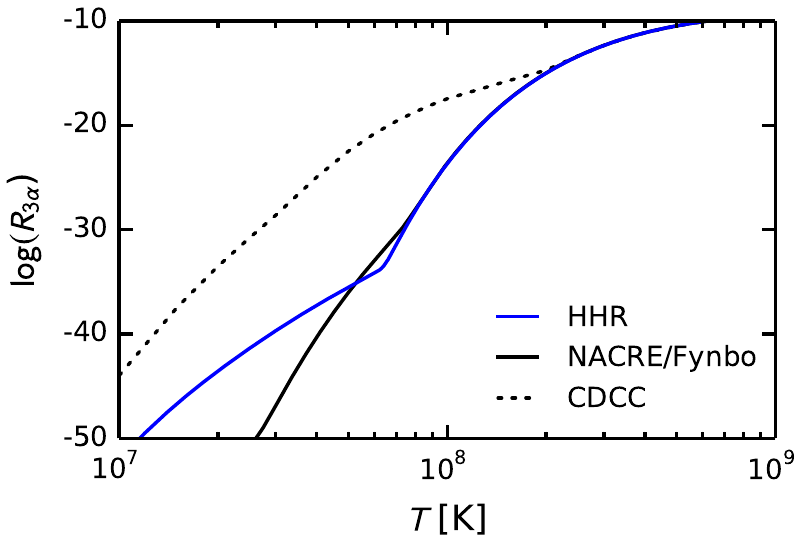}
\vspace{-0.4cm}
\caption{A comparison of the HHR \citep{Nguyen_2012}, NACRE \citep{Angulo_1999}, and CDCC \citep{Ogata_2009} 3$\alpha$ rates. At this scale the rate of \citet{Fynbo_2005} is indistinguishable from the NACRE rate.}
\label{fig:3arates}
\end{figure}

We examine helium novae for two 3$\alpha$ rates, the HHR rate and the NACRE rate, which are shown in Figure~\ref{fig:3arates} along with the rate by \citet{Ogata_2009} for comparison. Figure~\ref{fig:3a_difference} shows the normalized rates in the temperature range $(0.5\textrm{--}1.0) \times 10^{8} \, \K$ where HHR is less than NACRE. We also briefly examine the $3\alpha$ rate by \citet{Fynbo_2005}.

\section{Helium novae with the direct three-body rate} \label{sec:results}

We model pure $\helium$ accretion onto a CO WD with a mass $M = 1 \, \Msun$. Additional WD masses are not considered because the helium ignition mass is largely insensitive to the mass of the accretor's core \citep{Iben_1991}.  We examine two initial central temperatures for the WD, $\Tc = 4.0\ee{7} \, \K$ ($L_{\rm WD} \approx 0.1 \, \Lsun$) and $\Tc =  1.6\ee{7} \, \K$ ($L_{\rm WD} \approx 0.01 \, \Lsun$), typical of studies of accreting white dwarfs \citep{Woosley_Kasen_2011, Brooks_2015, Piersanti_2015}.

\subsection{Denegerate Donor} \label{sec:degen_methods}

For the case of a degenerate He WD donor, the time-averaged accretion rate decreases as the system evolves and mass transfer widens the binary. We approximate the accretion rate using a power law (\citealt{Bildsten_2006}; see their Fig. 1)
\begin{equation}
\Mdot = 0.03 \left(t + t_0 \right)^{-1} \, \Msunperyr \ , 
\label{eqn:mdot}
\end{equation}
where $t$ is the time in years and $t_0$ is chosen to set the initial accretion rate. For example, to set an initial accretion rate of $\Mdot = 10^{-6} \, \Msunperyr$ we choose $t_0 = 3\ee4 \, \mathrm{years}$.

\citet{Piersanti_2015} find that accretion rates upon contact may vary from $\approx (0.5\textrm{--}2.0) \times 10^{-6} \, \Msunperyr$ for double WD binaries. For the cooler model with a central temperature $T_c = 1.6 \times 10^7 \, {\K}$, we choose three initial accretion rates: $\Mdot = 1.0, 1.5,$ and $2.0\ee{-6} \, \Msunperyr$. For the warmer model with $T_c = 4.0 \times 10^7 \, \K$, we use the same three initial accretion rates and an additional rate $\Mdot = 0.5\ee{-6} \, \Msunperyr$. For each choice of the initial accretion rate, we run models using both the NACRE $3\alpha$ rate and the HHR $3\alpha$ rate. 

\begin{figure} 
\centering
\includegraphics[width=1.0\columnwidth]{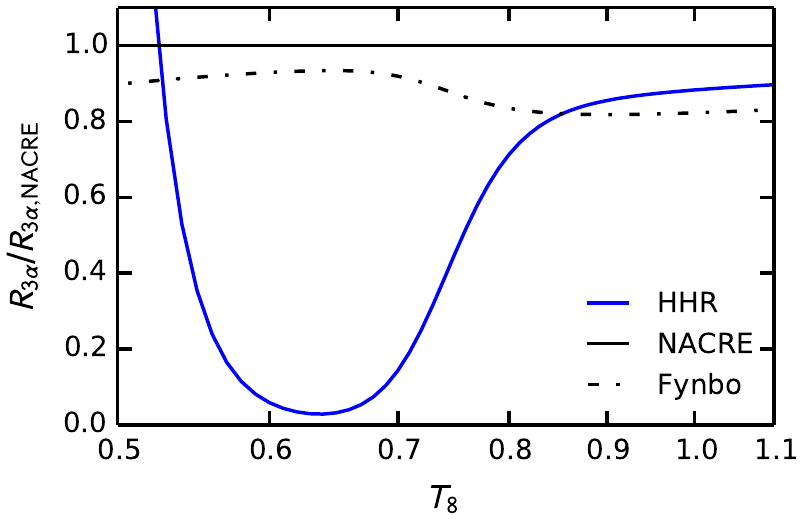}
\vspace{-0.4cm}
\caption{HHR $3\alpha$ rate (solid blue curve) normalized to the NACRE rate. The rate by \citet{Fynbo_2005} is included as well (dot-dashed black curve).}
\label{fig:3a_difference}
\end{figure}

\begin{figure} 
\centering
\includegraphics[width=1.0\columnwidth]{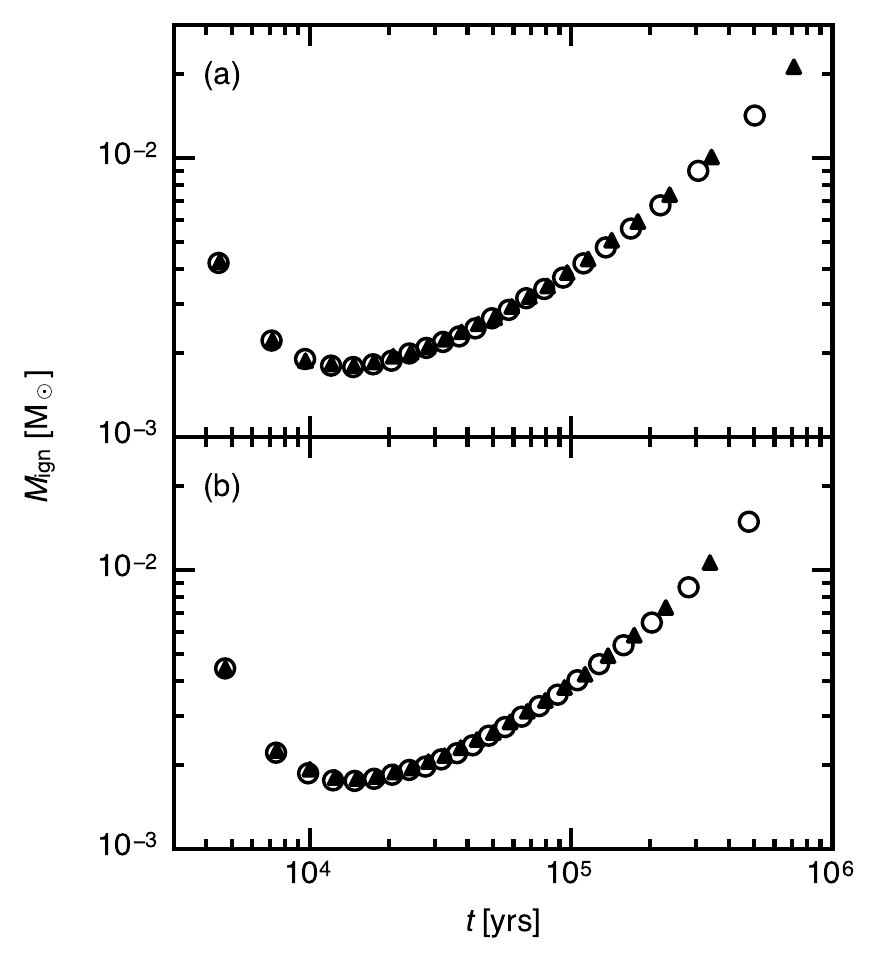}
\vspace{-0.4cm}
\caption{The envelope mass at ignition of helium burning, $\Mign$, through the duration of an evolutionary run with a $1 \, \Msun$ CO WD having initial $\Mdot = 10^{-6} \, \Msunperyr$. The open circles are models using the NACRE $3\alpha$ rate and the triangles are models using the HHR rate. {\it Panel (a):} WD with initial core temperature of $\Tc = 4.0 \times 10^{7} \, {\K}$. {\it Panel (b):} WD with initial core temperature of $\Tc = 1.6 \times 10^{7} \, {\K}$. }
\label{fig:Mign_1ab}
\end{figure}

As can be seen in the top panel of Figure~\ref{fig:Mign_1ab}, accretion first proceeds onto a cold envelope at $T_c=4.0\times 10^7 \, \K$ and unstable helium ignition requires $\Mign \approx 4 \times 10^{-3} \, \mathrm{M_{\odot}}$. The first nova heats the envelope and the next few novae require $\Mign < 4 \times 10^{-3} \, \mathrm{M_{\odot}}$. After the 4th or 5th nova, subsequent novae require an increasingly more massive helium-rich shell to reach unstable helium ignition conditions. 

Initially, both the NACRE and HHR $3\alpha$ rates require nearly identical values of $M_{\rm ign}$ to produce novae. As the models evolve, however, the novae using the HHR rate have increasingly longer recurrence times compared to the novae using the NACRE rate, and as a result the ignition masses using the HHR rate are larger. This is most evident in the final, most energetic nova, which occurs about $\approx 200,000 \, \mathrm{yrs}$ later using the HHR rate and with a $\sim$$50\%$ larger ignition mass compared to the NACRE case. One might expect that the HHR rate, which is greatly enhanced at temperatures $T \lesssim 5 \times 10^7 \, \K$, would ignite unstable helium burning sooner than the NACRE rate and thus lead to less energetic novae. We find the opposite: HHR novae have larger ignition masses and larger recurrence times compared to NACRE. This is because unstable helium ignition (when $\Lnuc \gtrsim L_\mathrm{WD}$) occurs in the accreted helium envelope between $T \approx (0.8 \textrm{--} 1.0) \times 10^8 \, \K$ where the HHR rate is reduced relative to NACRE, and therefore HHR requires a more massive helium layer to ignite unstable helium burning. 

We also test the $3\alpha$ rate of \citet{Fynbo_2005}. The Fynbo rate is lower than NACRE by $\approx 10\textrm{--}20 \%$ from $T \approx (0.8\textrm{--}1.0) \times 10^8 \, \K$, similar to the HHR rate. Figure~\ref{fig:3a_difference} shows the HHR and \citet{Fynbo_2005} rates normalized to the NACRE rate in this region of interest. We find that the Fynbo rate affects nova properties in the same way as HHR: later and more massive. 

The bottom panel of Figure \ref{fig:Mign_1ab} shows the results with the same initial accretion rate for a cooler WD with $\Tc = 1.6 \times 10^{7} \, {\K}$. Although the qualitative behavior remains the same \textrm{--} HHR results in consistently later and more massive novae \textrm{--} the last observed novae occur near $\approx 3 \times 10^5 \, \mathrm{yrs}$ for the HHR rate and near $\approx 5 \times 10^5 \, \mathrm{yrs}$ for the NACRE rate. The longer recurrence times with HHR mean that the accretion rate may drop too low for the envelope to acquire the necessary mass to ignite unstable helium burning before cooling significantly \citep{Shen_2009}.

\begin{figure} 
\centering
\includegraphics[width=1.0\columnwidth]{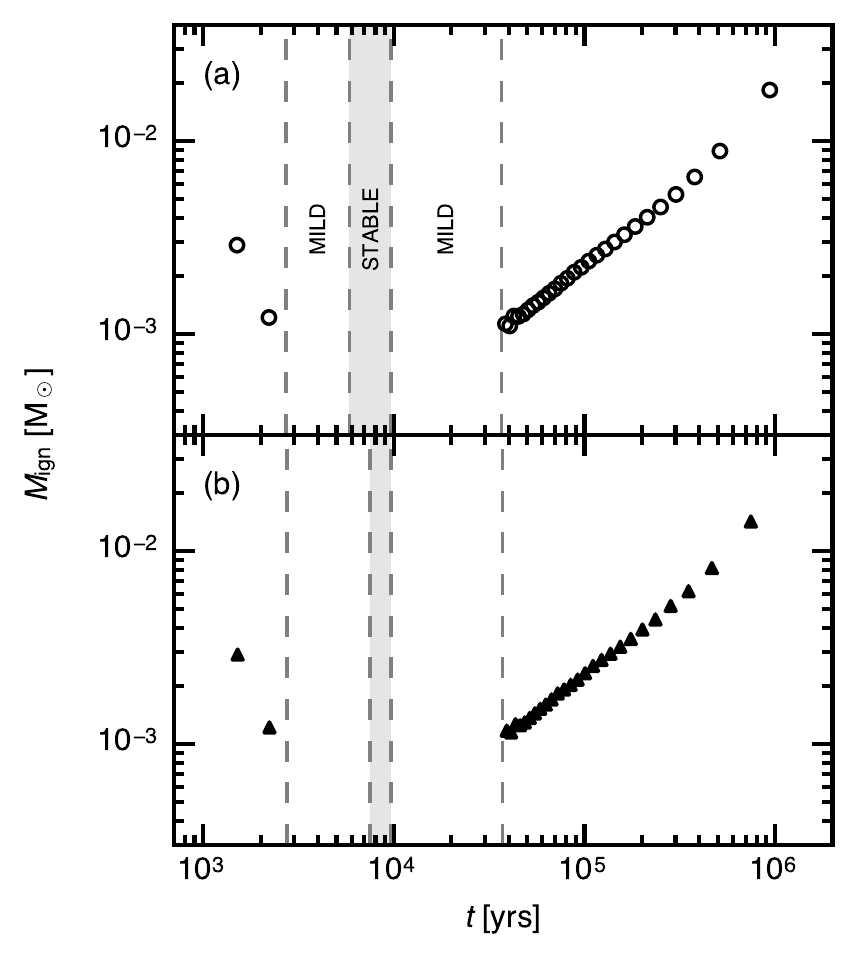}
\vspace{-0.4cm}
\caption{Envelope mass at ignition, $\Mign$, for a $1 \, \Msun$ CO WD with initial $\Mdot = 2 \times 10^{-6} \, \Msunperyr$ and $\Tc = 4.0 \times 10^{7} \, {\K}$. Dashed lines highlight the regions where novae become mild ($4 < \log L_\mathrm{nuc,peak} < 8$) and where burning is stable (shaded grey). {\it Panel (a):} results with the NACRE $3\alpha$ rate. {\it Panel (b):} results with the HHR rate.}
\label{fig:Mign_1_stable_regions}
\end{figure}

Cases where the initial accretion rate is larger than $\Mdot \gtrsim 10^{-6} \, \Msunperyr$ exhibit a period of stable burning. Figure~\ref{fig:Mign_1_stable_regions} illustrates this with the results from the models having an initial accretion rate of $\Mdot = 2\ee{-6} \, \Msunperyr$ onto the warmer $\Tc = 4.0\ee{7} \, \K$ WD. After the first few novae heat the initially cold envelope, novae become very mild (with $\Lnuc$ not exceeding $10^8 \, \Lsun$) and eventually helium burns stably. As the accretion rate decreases and approaches $\Mdot \sim 10^{-6} \, \Msunperyr$ mildly energetic novae resume and grow until the evolution of the system resembles the cases from Figure~\ref{fig:Mign_1ab} at late times.

\citet{Shen_2007} take an analytical approach to the stability of burning in the WD envelope. They find stable helium burning occurs for accretion rates $\gtrsim 2 \times 10^{-6} \, \mathrm{M_{\odot} \ yr^{-1}}$ for a $0.9 \, \Msun$ WD accreting pure helium. When following their derivation to calculate steady state burning conditions ($\rho,T$) for a broader range of white dwarf masses (0.4 to 1.35 $\Msun$) and accretion rates ($10^{-8}$ to $10^{-5.5} \, \Msunperyr$), we calculate that all of the steady state helium burning solutions exist above $T = 10^8 \, \K$, where the HHR and NACRE $3\alpha$ rates (and especially their derivatives with respect to temperature) are similar. With this approach, the accretion rates at which stable burning occurs does not change between the HHR and NACRE $3\alpha$ rates. In our numerical models, we find that stable burning episodes in models with higher initial accretion rates do behave slightly differently depending on the $3\alpha$ rate. With the HHR rate, the system experiences a few more mild flashes before transitioning into stable burning, with stable burning beginning about $\sim 10^3$ years later than the NACRE model. After stable burning, both rates transition to mild flashes and strong flashes at very similar times. The delayed transition from mild flashes to stable burning with the HHR rate emphasizes again the importance of the rate below $T \lesssim 10^8 \, \K$, where the analytical approach doesn't offer insight into the ignition of burning.

\subsection{Non-denegerate Donor} \label{sec:nondegen_methods}

\citet{Brooks_2015} examine the case of a low mass non-degenerate donor undergoing mass transfer via angular momentum loss due to gravitational wave radiation (rather than evolution of a higher mass donor; \citealt{Brooks_2016}). The first helium ignition has a more massive envelope than subsequent bursts, and could lead to a detonation of the helium layer. In such systems, the initial accretion rate at contact is near $\dot{M} \approx \mathrm{a \, few} \times 10^{-8} \, \Msunperyr$. The structural response of the helium star donor to mass loss alters the evolution of the binary. Once mass transfer begins, the binary first tightens, rather than widens. On the timescale of a few Myr, $\Mdot$ climbs to a maximum around $\sim 10^{-7} \, \Msunperyr$, corresponding to a binary period of about $\sim 10$ minutes. After reaching the period minimum, the binary begins to widen as a result of mass transfer, similar to the degenerate donor case. The accretion rate decreases monotonically until it becomes low enough that no further helium novae occur.

Because the accretion rate in the non-degenerate donor case starts low and onto a cold WD, the first nova often occurs with the largest helium envelope. Subsequent novae are smaller and more frequent as the accretion rate rises. Though the binary eventually widens and the accretion rate drops again, the later novae still tend to be weaker than the first because the accretion rate decreases too quickly after the period minimum to allow for many bursts. We impose a constant accretion rate of $\Mdot = 3.5\ee{-8} \, \Msunperyr$ onto the $1 \, \Msun$ WD with central temperature $T_c = 1.6 \times 10^7 \, {\K}$. We find that this choice of conditions results in a strong nova with about $0.08 \, \Msun$ of accreted helium envelope after accreting for just over $2 \, \Myr$, consistent with the result found by \citet{Brooks_2015}. 

As with the degenerate donor case, the HHR rate results in a later ignition with a more massive envelope. Since this is only a single accretion and nova event, however, the difference between the two $3\alpha$ rates is not as pronounced as in the degenerate donor model. We find that the differences due to the $3\alpha$ rates here is only a few percent in time of ignition and envelope mass. The timescales to reach the most energetic nova in the degenerate models are all less than $\approx 1 \, \mathrm{Myr}$ (after which accretion is too slow to heat the envelope), but the non-degenerate models ignite after $\gtrsim 2 \, \mathrm{Myr}$. Yet the difference in ignition time and envelope mass between $3\alpha$ rates is relatively small, compared to the larger compounded effect in the degenerate case. As in the degenerate donor case, we also test the Fynbo $3\alpha$ rate and find the same qualitative behavior as the HHR rate: ignition occurs later with a more massive helium layer.

\section{Discussion} \label{sec:conclusions}

In this work we examine the observational impact of the direct three-body 3$\alpha$ rate on helium novae in AM CVn systems. We find that the direct three-body HHR rate results in helium novae with more massive envelopes at unstable helium ignition compared to the NACRE rate. Furthermore, in systems with recurring novae, the HHR rate results in longer recurrence times between novae, as can be seen in Figure~\ref{fig:Mign_1ab}. The most dramatic differences in our tested cases show the final nova occurring $\sim 50\%$ later and with $\sim 50\%$ more massive helium envelopes using the HHR $3\alpha$ rate. Because unstable helium ignition occurs near $T \approx (0.8\textrm{--}1.0) \times 10^8 \K$, the reduction in the HHR rate relative to NACRE at temperatures $T \gtrsim 5 \times 10^7 \, \mathrm{K}$ primarily drives the differences in nova properties. Moreover, we find that the Fynbo rate \textrm{---} which is also reduced relative to NACRE between $T \approx (0.8\textrm{--}1.0) \times 10^8 \K$ \textrm{---} produces similar results to models using the HHR rate. 

In systems with a degenerate helium donor, the time and envelope mass of the final nova is sensitive to the initial accretion rate and core temperature, which are not well constrained. Novae in models using the HHR rate require a larger ignition mass than the NACRE rate and have increasingly longer recurrence times. As the binary evolves and the accretion rate declines, near $\Mdot \sim 10^{-8} \, \Msunperyr$ accretion becomes too slow to build up a sufficient envelope mass to ignite unstable helium burning. If the final ignition with the NACRE rate for a given set of initial conditions ($\Mdot$, $\Tc$) occurs sufficiently late, then there may be no corresponding ignition with the HHR rate. That is to say, in some cases both the HHR and NACRE rate exhibit $N$ helium ignitions, with the last HHR flash occurring later. In other cases, the HHR rate may only exhibit $N-1$ ignitions, with its last flash occurring earlier than NACRE. If the initial conditions of AM CVn systems with degenerate donors are sampled randomly in the range of parameter space studied here, then the effect of the $3\alpha$ rate on the final, most energetic event is essentially random as well. Unless we can better constrain (or understand the distribution of) the initial $\Mdot$ and $\Tc$ in these systems, we can not conclude if there will be a systematic observable effect of the HHR (or Fynbo) rate on the frequency of potential ``.Ia supernovae" events in future surveys. 

We also examine the case of a non-degenerate helium donor. The accretion rate in these systems is initially very low, and therefore the first helium nova often has a more massive envelope than subsequent novae \citep{Brooks_2015}. With the HHR $3\alpha$ rate we find that the first ignition event with the non-degenerate donor occurs $\approx 5\times 10^4 \, \mathrm{yrs}$ later and with a $\approx 2\%$ more massive envelope than with the NACRE rate. Because the quantitative difference in novae properties is very small compared to the final novae in systems with a degenerate donor, the differences in rates is seen more clearly over many recurrent novae. Even if the HHR rate consistently causes later and more energetic novae in systems with non-degenerate donors, changes at the few percent level in frequency and energetics of the novae will likely be difficult to detect observationally. 

\citet{Woosley_Kasen_2011} found that a critical density $\rho \gtrsim 10^6 \, \mathrm{g} \, \mathrm{cm}^{-3}$ at unstable helium ignition is necessary to cause a detonation of the helium layer. In our non-degenerate case, the largest accreted envelopes at ignition are near $\approx 0.08 \, \Msun$ and the peak densities are near $\rho \approx 8 \times 10^5 \, \mathrm{g} \, \mathrm{cm}^{-3}$ as the runaway occurs, with HHR and Fynbo reaching densities only $\approx 5\%$ larger than NACRE. Though this difference is small, the HHR and Fynbo $3\alpha$ rates serve to delay the onset of ignition to higher densities nonetheless. An increase in ignition density near the threshold for detonation of the helium layer due to changes in the $3\alpha$ rate should motivate future studies.

\acknowledgements
The authors thank Ken Shen for helpful discussions. This material is based upon work supported by the National Science Foundation under Grant No. AST-1516969 and Grant No. PHY-1430152 (Joint Institute for Nuclear Astrophysics \textrm{--} Center for the Evolution of the Elements).

\bibliographystyle{apj}
\bibliography{3a}

\begin{thebibliography}{30}
\expandafter\ifx\csname natexlab\endcsname\relax\def\natexlab#1{#1}\fi

\bibitem[{Angulo {et~al.}(1999)Angulo, Arnould, Rayet, Descouvemont, Baye,
  Leclercq-Willain, Coc, Barhoumi, Aguer, Rolfs, Kunz, Hammer, Mayer,
  Paradellis, Kossionides, Chronidou, Spyrou, Degl'Innocenti, Fiorentini,
  Ricci, Zavatarelli, Providencia, Wolters, Soares, Grama, Rahighi, Shotter, \&
  Rachti}]{Angulo_1999}
Angulo, C., {et~al.} 1999, \nphysa, 656, 3

\bibitem[{Bildsten {et~al.}(2007)Bildsten, Shen, Weinberg, \&
  Nelemans}]{Bildsten_2007}
Bildsten, L., Shen, K.~J., Weinberg, N.~N., \& Nelemans, G. 2007, \apjl, 662,
  L95

\bibitem[{Bildsten {et~al.}(2006)Bildsten, Townsley, Deloye, \&
  Nelemans}]{Bildsten_2006}
Bildsten, L., Townsley, D.~M., Deloye, C.~J., \& Nelemans, G. 2006, \apj, 640,
  466

\bibitem[{{Brooks} {et~al.}(2015){Brooks}, {Bildsten}, {Marchant}, \&
  {Paxton}}]{Brooks_2015}
{Brooks}, J., {Bildsten}, L., {Marchant}, P., \& {Paxton}, B. 2015, \apj, 807,
  74

\bibitem[{{Brooks} {et~al.}(2016){Brooks}, {Bildsten}, {Schwab}, \&
  {Paxton}}]{Brooks_2016}
{Brooks}, J., {Bildsten}, L., {Schwab}, J., \& {Paxton}, B. 2016, \apj, 821, 28

\bibitem[{{Deloye} \& {Bildsten}(2003)}]{Deloye_2003}
{Deloye}, C.~J., \& {Bildsten}, L. 2003, \apj, 598, 1217

\bibitem[{{Deloye} {et~al.}(2005){Deloye}, {Bildsten}, \&
  {Nelemans}}]{Deloye_2005}
{Deloye}, C.~J., {Bildsten}, L., \& {Nelemans}, G. 2005, \apj, 624, 934

\bibitem[{Denissenkov {et~al.}(2013)Denissenkov, Herwig, Bildsten, \&
  Paxton}]{Denissenkov_2013}
Denissenkov, P.~A., Herwig, F., Bildsten, L., \& Paxton, B. 2013, \apj, 762, 8

\bibitem[{Dotter \& Paxton(2009)}]{Dotter_2009}
Dotter, A., \& Paxton, B. 2009, \aap, 507, 1617

\bibitem[{Ferguson {et~al.}(2005)Ferguson, Alexander, Allard, Barman, Bodnarik,
  Hauschildt, Heffner-Wong, \& Tamanai}]{Ferguson_2005}
Ferguson, J.~W., Alexander, D.~R., Allard, F., Barman, T., Bodnarik, J.~G.,
  Hauschildt, P.~H., Heffner-Wong, A., \& Tamanai, A. 2005, \apj, 623, 585

\bibitem[{{Foley} {et~al.}(2009){Foley}, {Chornock}, {Filippenko},
  {Ganeshalingam}, {Kirshner}, {Li}, {Cenko}, {Challis}, {Friedman}, {Modjaz},
  {Silverman}, \& {Wood-Vasey}}]{2009AJ....138..376F}
{Foley}, R.~J., {et~al.} 2009, \aj, 138, 376

\bibitem[{Fynbo {et~al.}(2005)Fynbo, Diget, Bergmann, Borge, Cederkäll,
  Dendooven, Fraile, Franchoo, Fedosseev, Fulton, Huang, Huikari, Jeppesen,
  Jokinen, Jones, Jonson, Köster, Langanke, Meister, Nilsson, Nyman, Prezado,
  Riisager, Rinta-Antila, Tengblad, Turrion, Wang, Weissman, Wilhelmsen, \&
  Äystö}]{Fynbo_2005}
Fynbo, H. O.~U., {et~al.} 2005, \nat, 433, 136

\bibitem[{Hoyle(1954)}]{Hoyle_1954}
Hoyle, F. 1954, \apjs, 1, 121

\bibitem[{{Iben} \& {Tutukov}(1991)}]{Iben_1991}
{Iben}, Jr., I., \& {Tutukov}, A.~V. 1991, \apj, 370, 615

\bibitem[{Iglesias \& Rogers(1993)}]{Iglesias_1993}
Iglesias, C.~A., \& Rogers, F.~J. 1993, \apjl, 412, 752

\bibitem[{Iglesias \& Rogers(1996)}]{Iglesias_1996}
---. 1996, \apjl, 464, 943

\bibitem[{{Kasliwal} {et~al.}(2008){Kasliwal}, {Ofek}, {Gal-Yam}, {Rau},
  {Brown}, {Cenko}, {Cameron}, {Quimby}, {Kulkarni}, {Bildsten}, {Milne}, \&
  {Bryngelson}}]{2008ApJ...683L..29K}
{Kasliwal}, M.~M., {et~al.} 2008, \apjl, 683, L29

\bibitem[{Kasliwal {et~al.}(2010)Kasliwal, Kulkarni, Gal-Yam, Yaron, Quimby,
  Ofek, Nugent, Poznanski, Jacobsen, Sternberg, Arcavi, Howell, Sullivan, Rich,
  Burke, Brimacombe, Milisavljevic, Fesen, Bildsten, Shen, Cenko, Bloom, Hsiao,
  Law, Gehrels, Immler, Dekany, Rahmer, Hale, Smith, Zolkower, Velur, Walters,
  Henning, Bui, \& McKenna}]{Kasliwal_2010}
Kasliwal, M.~M., {et~al.} 2010, \apj, 723, L98

\bibitem[{Nguyen {et~al.}(2012)Nguyen, Nunes, Thompson, \& Brown}]{Nguyen_2012}
Nguyen, N.~B., Nunes, F.~M., Thompson, I.~J., \& Brown, E.~F. 2012, \prl, 109

\bibitem[{Ogata {et~al.}(2009)Ogata, Kan, \& Kamimura}]{Ogata_2009}
Ogata, K., Kan, M., \& Kamimura, M. 2009, Progress of Theoretical Physics, 122,
  1055

\bibitem[{{Paxton} {et~al.}(2011){Paxton}, {Bildsten}, {Dotter}, {Herwig},
  {Lesaffre}, \& {Timmes}}]{Paxton_2011}
{Paxton}, B., {Bildsten}, L., {Dotter}, A., {Herwig}, F., {Lesaffre}, P., \&
  {Timmes}, F. 2011, \apjs, 192, 3

\bibitem[{{Paxton} {et~al.}(2013){Paxton}, , Cantiello, Arras, Bildsten, Brown,
  Dotter, Mankovich, Montgomery, Stello, Timmes, \& Townsend}]{Paxton_2013}
{Paxton}, B., {et~al.} 2013, \apjs, 208, 4

\bibitem[{{Paxton} {et~al.}(2015){Paxton}, {Marchant}, {Schwab}, {Bauer},
  {Bildsten}, {Cantiello}, {Dessart}, {Farmer}, {Hu}, {Langer}, {Townsend},
  {Townsley}, \& {Timmes}}]{Paxton_2015}
---. 2015, \apjs, 220, 15

\bibitem[{{Piersanti} {et~al.}(2015){Piersanti}, {Yungelson}, \&
  {Tornamb{\'e}}}]{Piersanti_2015}
{Piersanti}, L., {Yungelson}, L.~R., \& {Tornamb{\'e}}, A. 2015, \mnras, 452,
  2897

\bibitem[{{Poznanski} {et~al.}(2010){Poznanski}, {Chornock}, {Nugent}, {Bloom},
  {Filippenko}, {Ganeshalingam}, {Leonard}, {Li}, \& {Thomas}}]{Poznanski_2010}
{Poznanski}, D., {et~al.} 2010, Science, 327, 58

\bibitem[{{Shen}(2015)}]{2015ApJ...805L...6S}
{Shen}, K.~J. 2015, \apjl, 805, L6

\bibitem[{Shen \& Bildsten(2007)}]{Shen_2007}
Shen, K.~J., \& Bildsten, L. 2007, \apj, 660, 1444

\bibitem[{{Shen} \& {Bildsten}(2009)}]{Shen_2009}
{Shen}, K.~J., \& {Bildsten}, L. 2009, \apj, 699, 1365

\bibitem[{{Warner}(1995)}]{Warner_1995}
{Warner}, B. 1995, \apss, 225, 249

\bibitem[{{Woosley} \& {Kasen}(2011)}]{Woosley_Kasen_2011}
{Woosley}, S.~E., \& {Kasen}, D. 2011, \apj, 734, 38

\end{thebibliography}


\end{document}